\documentclass[10pt]{iopart}

\usepackage{graphicx}
\usepackage{subcaption}
\usepackage{amssymb}
\usepackage[utf8]{inputenc}
\usepackage[english]{babel}
\usepackage[T1,T2A]{fontenc} 

\begin{document}

\title[Positive streamers in airflow]{DC-driven positive streamer coronas in airflow}

\author{Benjamin C. Martell, Lee R. Strobel, and Carmen Guerra-Garcia}

\address{Massachusetts Institute of Technology, 77 Massachusetts Ave, Cambridge, MA 02139, USA}
\ead{martellb@mit.edu, guerrac@mit.edu}
\vspace{10pt}

\begin{abstract}
An experimental study of the effect of laminar airflow on positive self-pulsating streamer coronas in a needle-to-plate geometry is presented. The experiments are performed in an open return wind tunnel with winds up to 30 m/s orthogonal to the needle. The experimental data is presented in terms of statistical properties of the discharge, inferred from high resolution, large sample-size current waveforms. The key properties of the current pulsations, namely inter-pulse period, peak current, deposited energy, and pulse width are analyzed as a function of wind speed and applied DC voltage. All parameters increase in dispersion with wind speed. The mean of the inter-pulse period decreases with wind speed and the mean pulsation frequency increases. The peak currents and energies per pulsation have a general tendency to decrease in magnitude but also higher-current, higher-energy, streamer bursts are observed. At low wind speeds, streamers preferentially propagate in the downwind direction but, as the wind speed is increased, more streamers can propagate upwind. Synchronized imaging reveals correlations between the streamer direction and the electrical characteristics.

\end{abstract}

%
\vspace{2pc}

\submitto{\PSST}
\vspace{2pc}
%
%
\ioptwocol

\section{Introduction}
Recent numerical studies have captured the effects of airflow on positive \cite{niknezhad_three-dimensional_2021} and negative \cite{niknezhad_dynamics_2021} streamer discharges, when exposed to lateral wind. Streamers are filamentary ionized structures that propagate at speeds of $10^5$ to $10^6$ m/s for $10^{-8}$ to $10^{-7}$ s in atmospheric pressure air \cite{briels_positive_2008,ebert_multiscale_2006,luque_positive_2008}, leaving behind an ionized channel. Whereas realistic wind speeds can have no influence at the timescales of streamer propagation, they will impact the behavior of the discharge at the longer timescales of ion motion by enhancing charge transport. In addition, flow patterns, such as recirculation bubbles or turbulent eddies, can temporarily trap or move ions, modifying the electric fields. The recent numerical results by Niknezhad et al. \cite{niknezhad_three-dimensional_2021} reveal that the long-lived ($10^{-5}$ s) ionization trail of a single positive streamer exposed to wind tilts in the direction of the wind, while remaining attached to the anode. The simulation also found that a successive streamer will follow the pre-ionized path of the previous streamer, adopting a curved shape in the direction of the wind. Qualitatively, this behavior has been experimentally observed by Vogel et al. \cite{Vogel2018ExperimentalFlow}. For negative Trichel corona discharges, Guo et al. \cite{guo_effect_2021} found that the filamentary discharge angle with respect to the vertical increases with wind speed (tilting towards the wind direction) and decreases with applied voltage (aligning with the main vertical electric field), using both experiments and numerical models.

Whereas these results present a leap in our understanding of the physics of streamers in wind, they are limited in that they consider single streamer behavior \cite{nijdam_physics_2020,bagheri_comparison_2018,pavan_investigations_2020}, which is rarely observed in practice. More often than not, streamers appear in ensembles or streamer coronas, a phenomenon that can not yet easily be addressed using numerical models \cite{luque_growing_2014} and requires experimental treatment. Moreover, the streamer bursts can appear superimposed to an underlying glow corona, which contributes to the space charge shielding \cite{Loeb1965,Goldman1978}. For streamer coronas driven by DC voltage (so-called onset streamers or transient coronas), the effect of wind is of particular interest since space charge transport is thought to be the key factor driving the self-pulsating behavior of the discharge \cite{Loeb1965,Goldman1978}. The influence of airflow in self-pulsating discharges driven by DC-voltage has implications in a wide range of problems ranging from atmospheric electricity \cite{arcanjo_observations_2021} to plasma sources for chemical treatment or bio-engineering \cite{wang_dc-driven_2017,khun_various_2018,janda_kinetic_2018}.

In this paper, we present wind tunnel experiments of self-pulsating positive streamer coronas under DC voltage in a tip-to-plate geometry and characterize the emergence of flow-induced regimes. We investigate the effect of applied voltage and wind speed on the electrical properties and structure of the discharge; and present a novel statistical treatment of key properties, including the frequency of pulsation, the amplitude of the current pulses, and the morphology and orientation of the streamer bursts.

\section{Experimental setup}

\begin{figure}[hbp]
    \centering
        \includegraphics[width=.5\textwidth,trim={0 0 0 0}, clip]{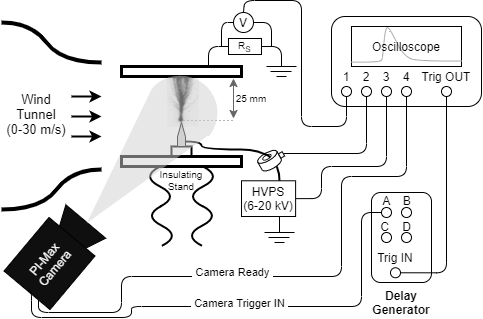}
    \caption{Schematic of the experimental setup.}
    \label{fig:exp_schematic}
\end{figure}

The influence of airflow on the self-pulsating positive streamer corona discharge is studied in a tip-to-plate geometry, exposed to lateral wind (relative to the needle orientation) as pictured in Figure \ref{fig:exp_schematic}.

The electrode geometry consists of a stainless steel needle-type electrode, with radius of curvature at the tip of around 100 $\mu$m, and an aluminum ground plate (with 203 mm diameter) located 25 mm from the tip of the needle. A Matsusada RB60-30 DC high voltage power supply is used to vary the applied voltage to the tip from 0 to +30 kV. 

The imaging studies presented from Figure~\ref{fig:allSuperimposed} onward use a slightly different tip made of tungsten and with a comparable radius of curvature. This modification was prompted by observed corrosion of the stainless steel needle induced by the discharge after several cycles of operation, but should not alter the results significantly (see supplementary material). Note that tip deterioration is a common issue when operating corona discharge reactors over several hours, and has been observed in both stainless steel and tungsten needles \cite{johnson_recent_2017,Ny1982CorrosionAir.,goldman_corona_1985}. For these studies, the tip was thoroughly cleaned with acetone between tests.  

The electrode setup is placed in the 46x46 cm test section of an open-return wind tunnel that provides laminar air speeds up to 30 m/s.

The 2 m long unshielded high voltage cable connecting the power supply to the needle passes through a rogowski coil (Pearson model 2877) that provides time-resolved measurement of the current at the high-voltage end. The current at the grounded end is measured using a resistor of $R_s$ = 180 $\mathrm{\Omega}$. Similar current traces are measured by both probes: the plots included in the paper correspond to the measurements taken by the ground-side probe and equivalent plots for the high-voltage end are provided in the supplementary material. Due to the moderate wind speeds tested, little to no leakage current in the direction of the flow is expected and the large (compared to the discharge gap) grounded plate captures most of the current, see~\ref{appendix} (Figure~\ref{fig:appendix4}). The power supply voltage is recorded for every test using the internal monitor of the device (not time resolved), however the high voltage applied to the needle was verified to be time-independent by using a time-resolved high voltage probe (Lecroy PPE 20kV). 

All electrical signals are measured by a Teledyne Lecroy Waverunner 9254 oscilloscope (4 GHz bandwidth). The data acquisition rate is  set to $50\cdot10^6$ Hz, and signals are recorded for 200 ms, which corresponds to data points acquired every 20 ns. These settings enable a large sampling of consecutive streamer bursts with adequate resolution. For example, a typical pulsation frequency of around 4 kHz would result in $\sim$800 streamer bursts sampled in each test.

\begin{figure}[htbp] 
    \centering
    \includegraphics[width=.48\textwidth,trim={0 0 0 0},clip]{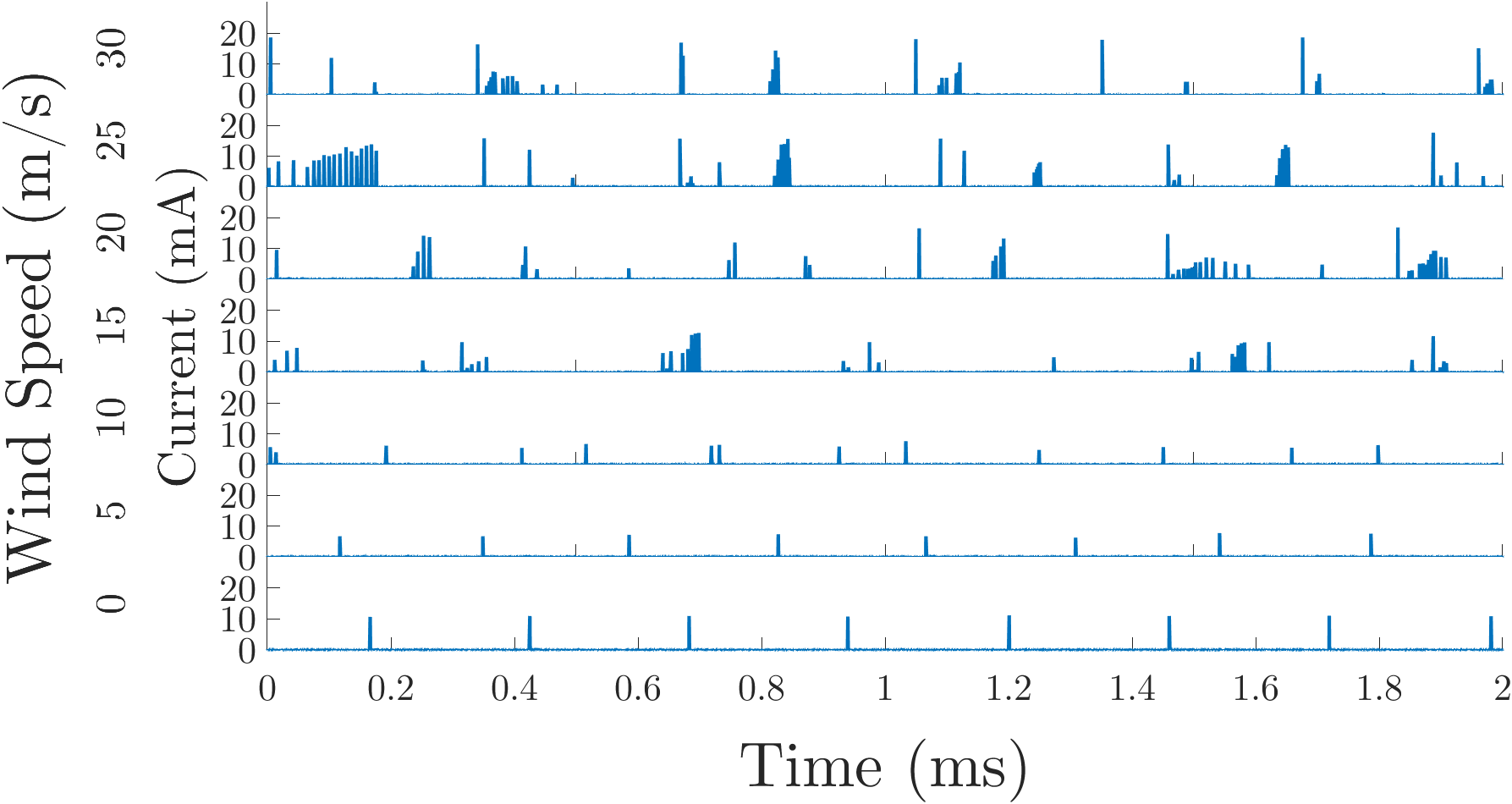}
    \caption{Current time-traces for an applied voltage of 16 kV and over the range of wind speeds tested.}
    \label{fig:raw_waveforms}
\end{figure}

Figure~\ref{fig:raw_waveforms} shows characteristic time traces of the current for wind speeds from 0 to 30 m/s and fixed applied voltage of 16 kV. Individual current pulses are identified for further analysis, as described in section~\ref{section:stat}.

\begin{figure}[htbp]
        \includegraphics[width=.48\textwidth,trim={0 0 0 0},clip]{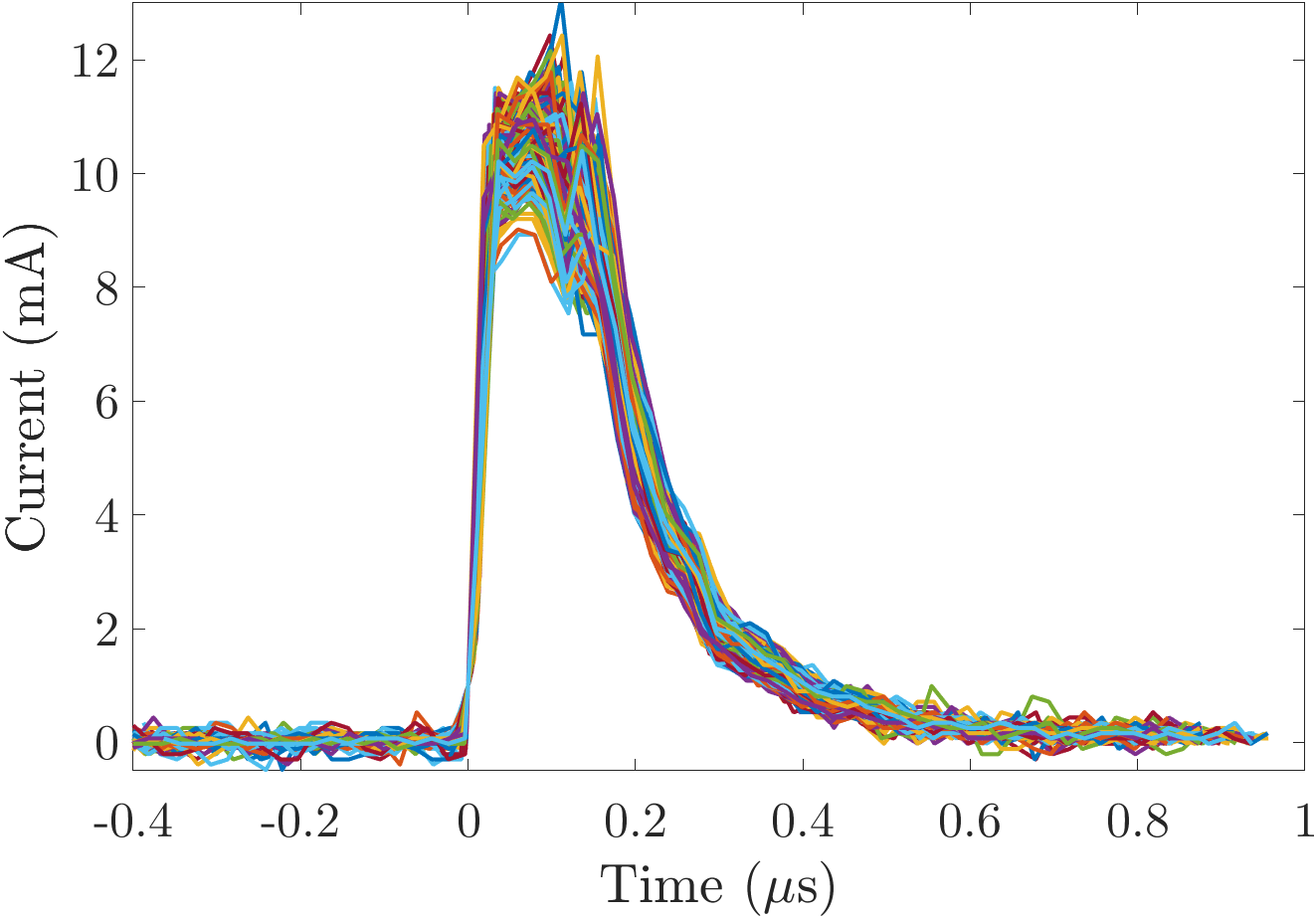} 
    \caption{Overlaid current pulses for all 760 streamer bursts in a given test. Case of no wind and 16 kV.}
    \label{fig:overlaid_waveforms}
\end{figure}

Figure~\ref{fig:overlaid_waveforms} shows 760 streamer bursts with an applied voltage of 16 kV and no wind, overlaid in time to visualize the pulse shape. It can be appreciated that, for this no-wind case, the pulses present little variation and have a repeatable pulse shape and amplitude, with little dispersion. The characteristic time scales of the pulse are consistent with prior works: rise times of tens of nanoseconds and decay times of hundreds of nanoseconds \cite{arcanjo_observations_2021}. 

\begin{figure}[htbp] 
    \subfloat[]{
        \includegraphics[width=.24\textwidth,trim={0 0 0 0},clip]{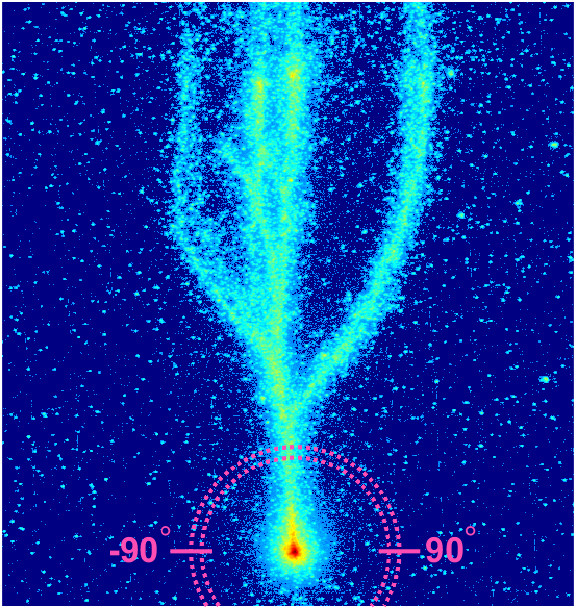} 
    } 
    \subfloat[]{
        \includegraphics[width=.24\textwidth,trim={0 0 0 0},clip]{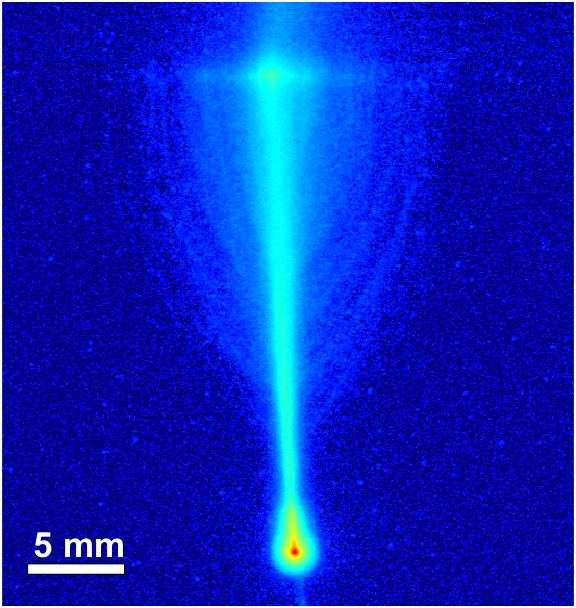} 
    } 
    \caption{(a) Single streamer burst and (b) multiple streamer bursts visualization using ICCD camera. Also shown is the angular parameter used to determine the directionality of the streamer bursts, from -90$^\circ$ to 90$^\circ$ (wind going to the right). Case of no wind and 16 kV.}
    \label{fig_DiagnosticsExample}
\end{figure}

Individual streamer corona bursts are imaged using a Princeton Instruments PI-MAX 4 Intensified Charged Coupled Device (ICCD) camera with a macro zoom lens (Vivitar Series 1 28-90mm) and a 20 mm macro extension to increase the image size on the detector. Images are taken with gate widths varying from 25 to 450 $\mathrm{\mu}$s (selected to be shorter than twice the inter-pulse period at those conditions) and maximum gain to capture the low luminosity of individual streamer bursts. Figure~\ref{fig_DiagnosticsExample}(a) shows a sample image of a single streamer burst, and Figure~\ref{fig_DiagnosticsExample}(b) shows the superposition of 152 of these images, highlighting the difference between the morphology of an individual burst and the time-integrated visualization. 

Two parameters are explored in this work: the applied voltage and the wind speed. Aside from those parameters, the measurements can also be influenced by the electrical circuit at hand, which is kept unmodified between tests. In particular, it is noted that the electrical circuit parameters can influence the pulse shape, e.g., through both the capacitance values of the discharging gap, as well as the high voltage cable and power supply \cite{janda_transient_2011}. The electrical circuit parameters can also affect the frequency of pulsations: e.g., increasing the value of a ballast resistance (placed between the power supply and the electrode) can decrease the frequency of pulsations \cite{wu_positive_2018, janda_transient_2011}, the current magnitude of the pulses, and the discharge regime at a given voltage and gap distance \cite{wu_positive_2018}.

\subsection{Statistical treatment of the electrical measurements}\label{section:stat}

Compared to streamer coronas triggered by pulsed nanosecond voltage \cite{briels_positive_2008,nijdam_experimental_2011}, self-pulsating streamer bursts triggered by DC voltage present more dispersion in their properties. In this work, this variability is encountered within a given applied voltage and wind-speed. The results in section~\ref{section:results} present a novel visualization of the streamer burst electrical properties, in terms of distribution functions rather than single-values.

To perform this statistical analysis, the acquisition of the current time traces needs to have high resolution and bandwidth (to resolve individual pulses) as well as high acquisition time (to sample large numbers of consecutive streamers and have statistically significant measurements). Considering these two competing criteria, the sampling rate was set to 50 MHz and the recording time at 200 ms. 

Analysis of the current time traces measured for any given voltage-wind speed pair begins by identifying the individual streamer bursts, which is done using a peak-finding algorithm. Once the peaks have been identified (typically around $\sim$800 for any given test) several parameters are measured. These include: the period of time in between consecutive streamer bursts, measured as the time from one peak to the next; the peak current magnitude, measured as the maximum value of the current pulse; the energy deposition in the gas, measured as the time-integral of the current-voltage product; and the width of the pulse, measured as the full-width at half-maximum (FWHM) of the current pulse. 
Depending on the applied voltage and wind speed condition, these measurements present more or less dispersion, and the results are plotted in terms of distribution functions, e.g., see Figures \ref{fig:period_histograms}-\ref{fig:width_histograms}. For each case, the darker histogram is the actual binned data, scaled relative to the global maximum of the full data set, so that all bins have equal weight. The lighter histograms are smoothed distributions and are all normalized to have the same maximum bin length for each distribution, to facilitate the visualization of the shape of the curve for regions that have fewer streamer bursts.

\subsection{Synchronization of streamer burst images to electrical signals} \label{section:synchronized}

Synchronized imaging of individual streamer bursts to the current time-traces is hindered by the non-negligible dispersion in the inter-pulse period. To achieve this synchronization, imaging of a streamer burst is triggered using the current pulse from the immediately preceding burst. To that end, a digital delay generator (Berkeley Nucleonics Corp. Model 577-4C) is used, which triggers signal acquisition a few microseconds after the oscilloscope is triggered by a simultaneous camera ready signal and a current pulsation. The gate width of the camera is selected to capture a single streamer burst and the current time-trace is saved for three consecutive streamer bursts: the trigger, the imaged one, and the subsequent. Since the camera is triggered by the preceding pulse, we limit the camera gate-width to less than twice the minimum inter-pulse period to avoid multiple streamers per image. Since the period is variable, and not known a priori, this method can filter out some of the streamer bursts with time from trigger to imaged burst greater than the camera gate width selected. To ensure that only one burst has been imaged, images corresponding to current time traces that present more than one current pulsation within the gate-width of the camera are discarded. This in turn limits the period after the pulse to be greater than the remaining gate-open time. These limitations especially affect the high wind speed cases in which the dispersion in the inter-pulse period is large. This data is presented in Section \ref{section:correlations} and Figures \ref{fig:period_after_correlations}-\ref{fig:periodCurrent}.

\section{Results}\label{section:results}

\subsection{Statistical treatment of current waveforms}

\begin{figure*}[htbp] 
        {\includegraphics[width=\textwidth,trim={0 0 0 0}, clip]{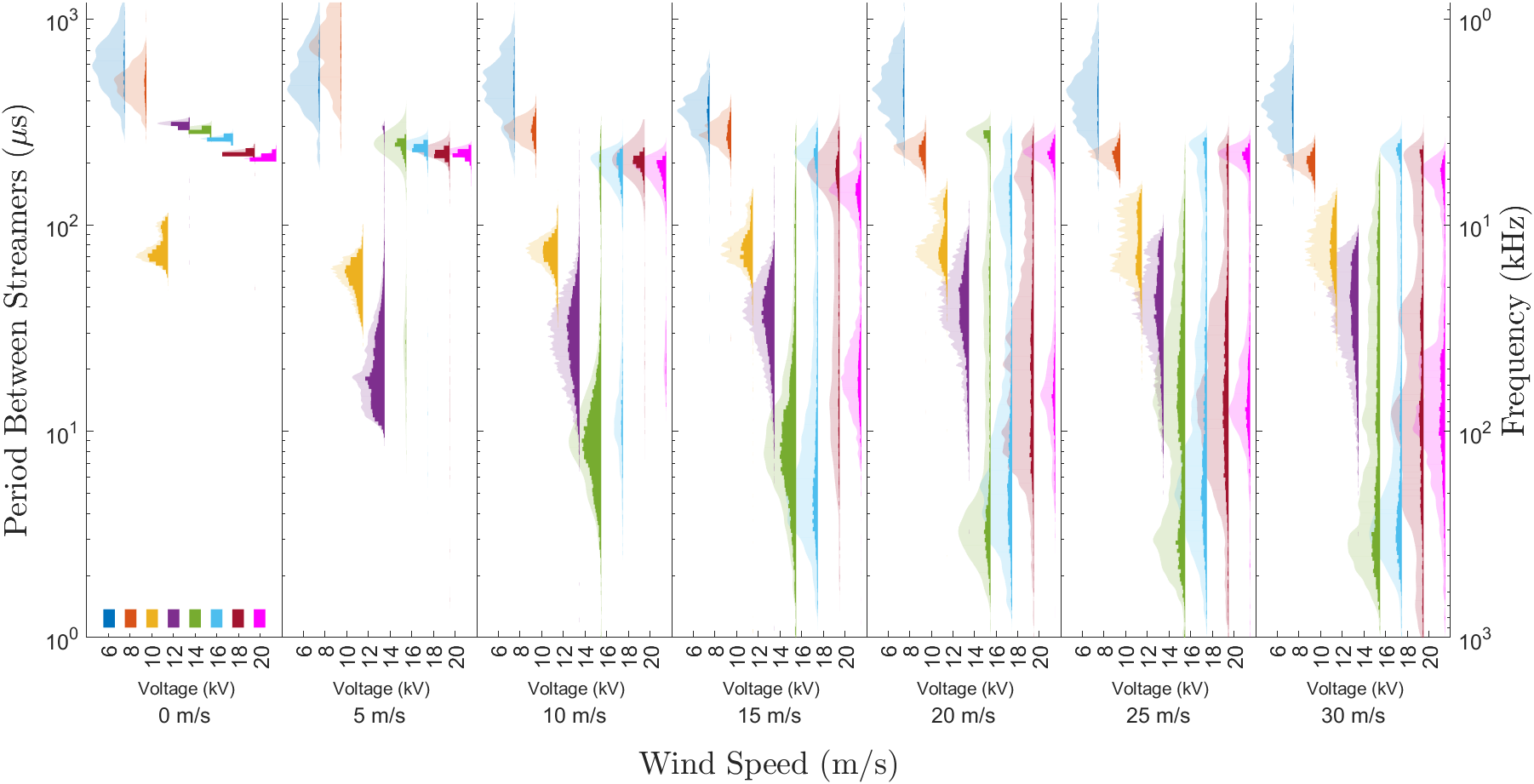}
    }
    \caption{Influence of applied voltage and wind speed on the inter-pulse period and pulse repetition frequency, presented as distribution functions. The darker histograms are the actual binned data. The lighter histograms are smoothed and normalized to have the same maximum.}\label{fig:period_histograms}
\end{figure*}

\begin{figure*}[htbp]    
    {\includegraphics[width=\textwidth,trim={0 0 0 0}, clip]{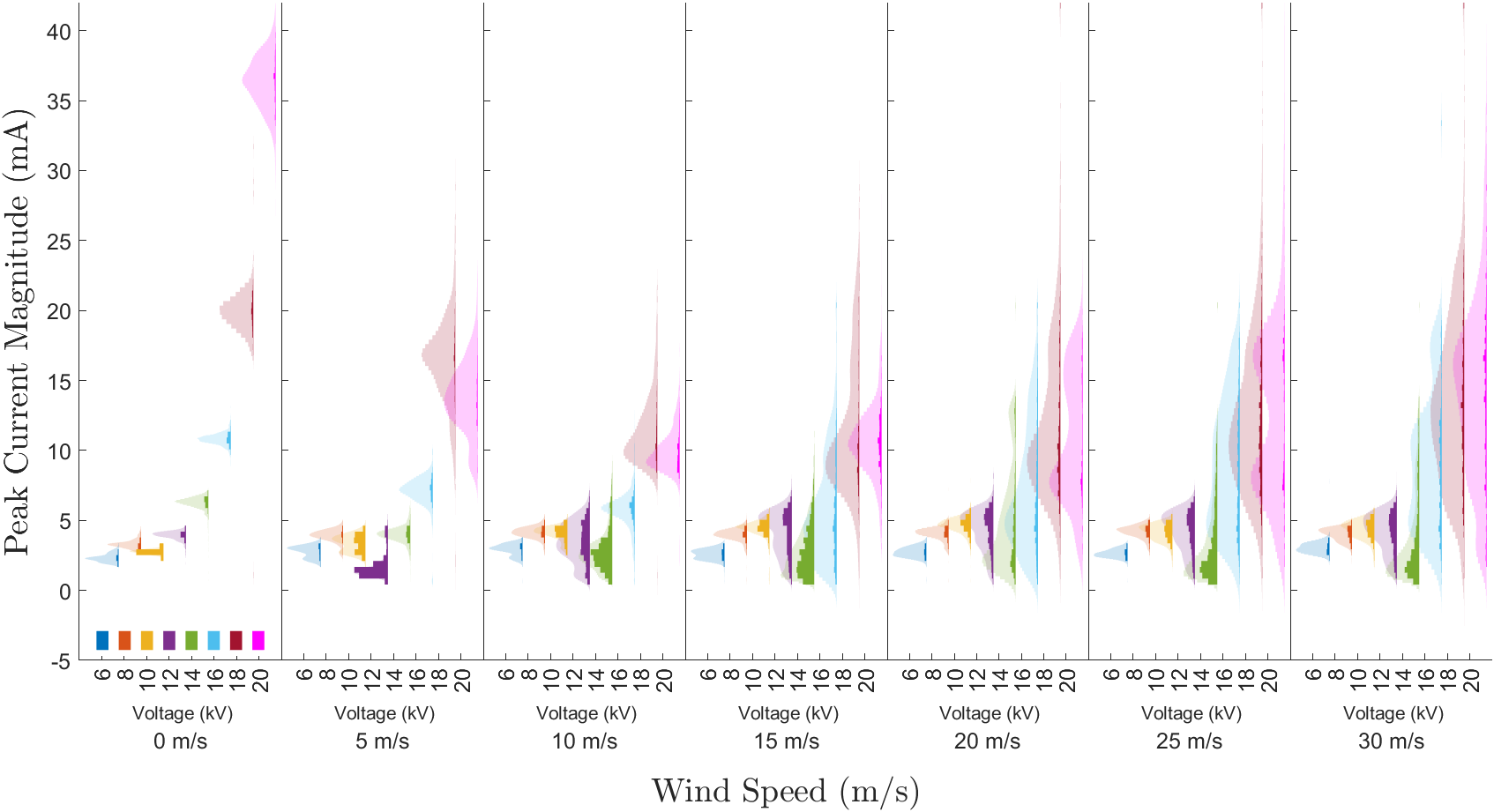}
    }
     \caption{Influence of applied voltage and wind speed on the peak current magnitude, presented as distribution functions. The darker histograms are the actual binned data. The lighter histograms are smoothed and normalized to have the same maximum.}\label{fig:peak_histograms}
\end{figure*}

\begin{figure*}[htbp] 
    {\includegraphics[width=\textwidth,trim={0 0 0 0}, clip]{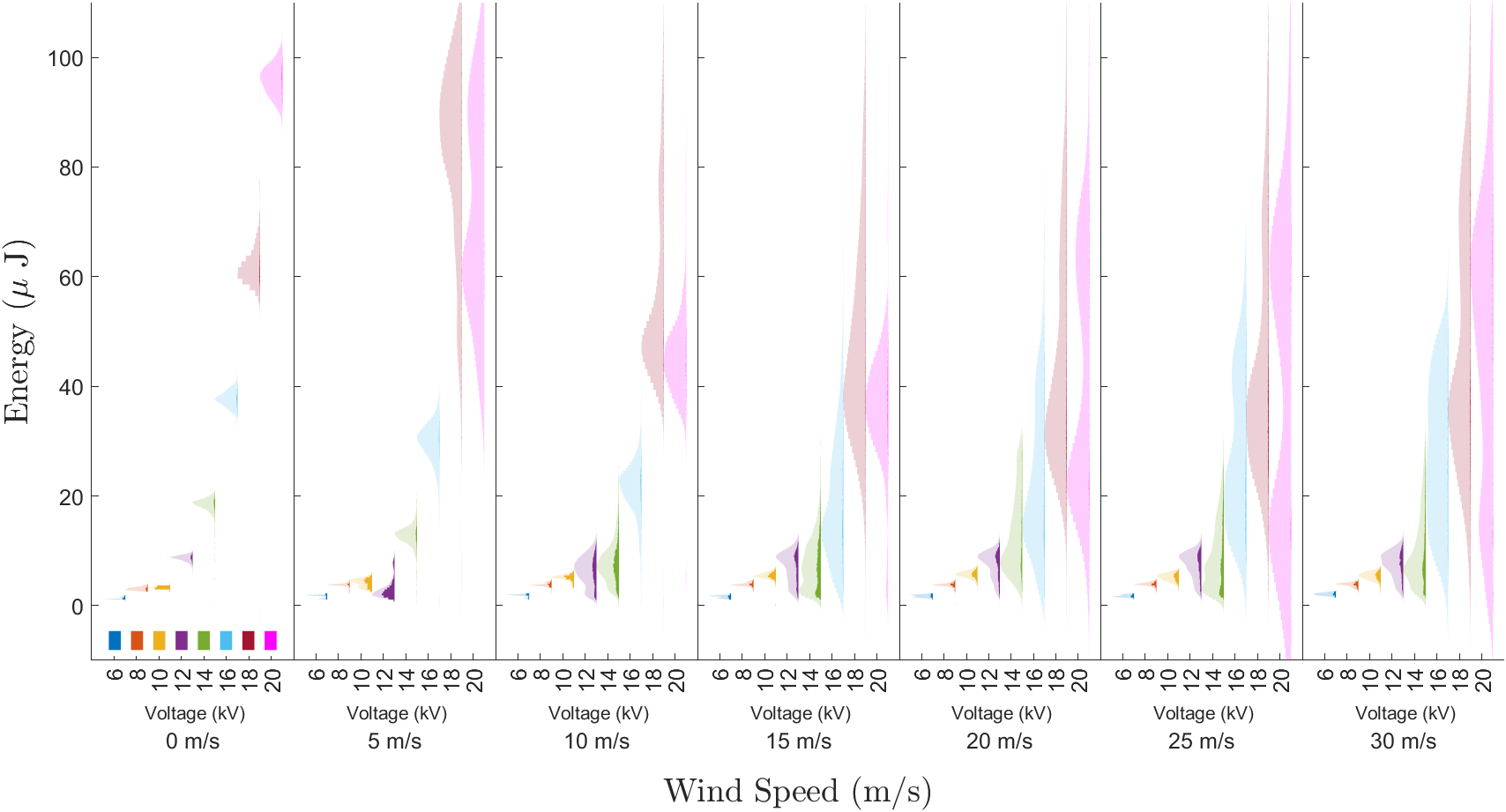}
    }
    \caption{Influence of applied voltage and wind speed on the energy deposition in the gas, presented as distribution functions. The darker histograms are the actual binned data. The lighter histograms are smoothed and normalized to have the same maximum.}\label{fig:energy_histograms}
\end{figure*}

\begin{figure*}[htbp]     
    {\includegraphics[width=\textwidth,trim={0 0 0 0}, clip]{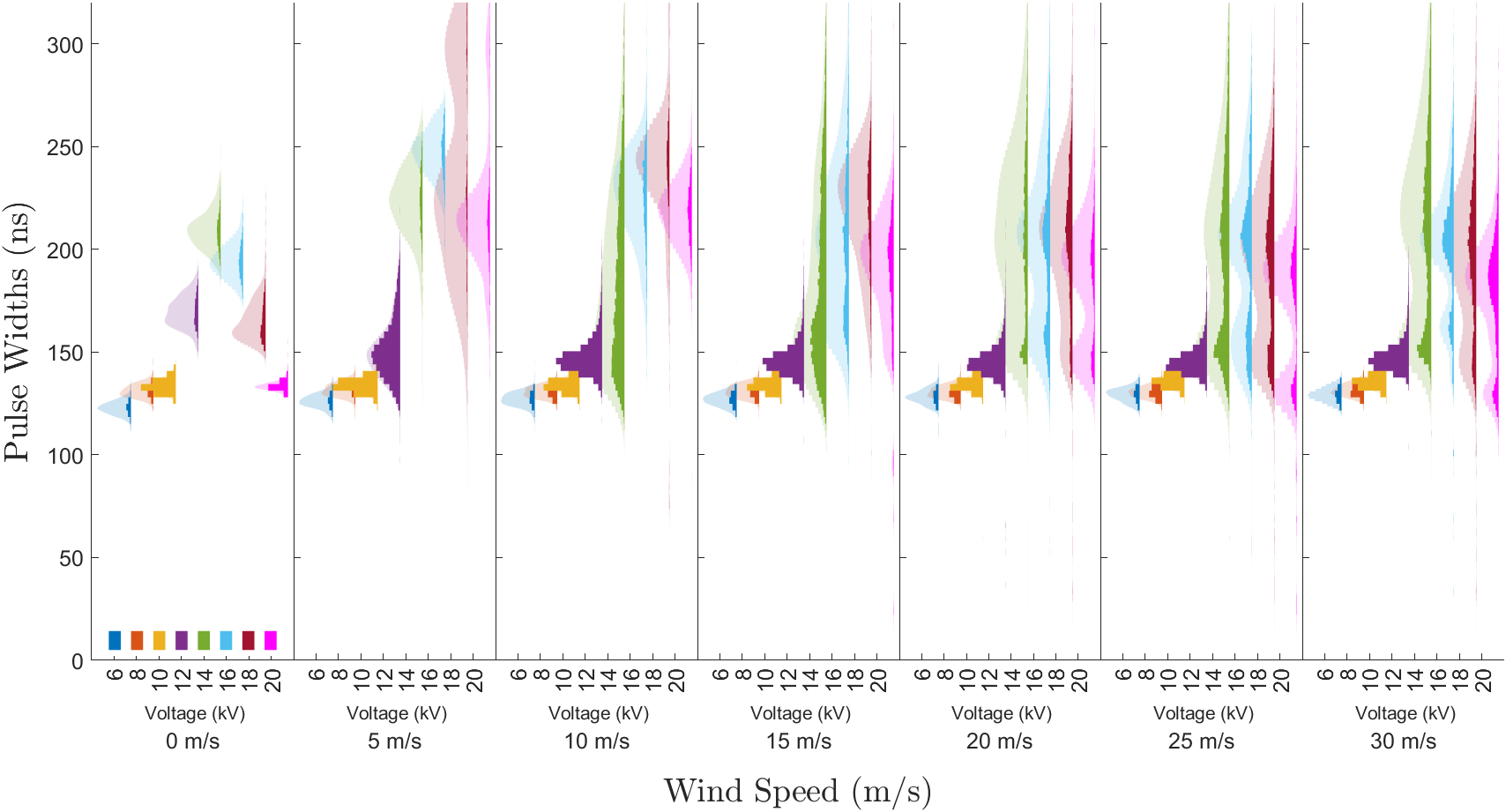}
    } 
    \caption{Influence of applied voltage and wind speed on the pulse width, presented as distribution functions. The darker histograms are the actual binned data. The lighter histograms are smoothed and normalized to have the same maximum.}\label{fig:width_histograms}
\end{figure*}

Figures~\ref{fig:period_histograms} through \ref{fig:width_histograms} present the distributions of the electrical properties of individual streamer pulsations as a function of applied voltage and wind speed.

\begin{figure*}
    \centering
        \includegraphics[width=\textwidth,trim={0 .5cm 0 0}, clip]{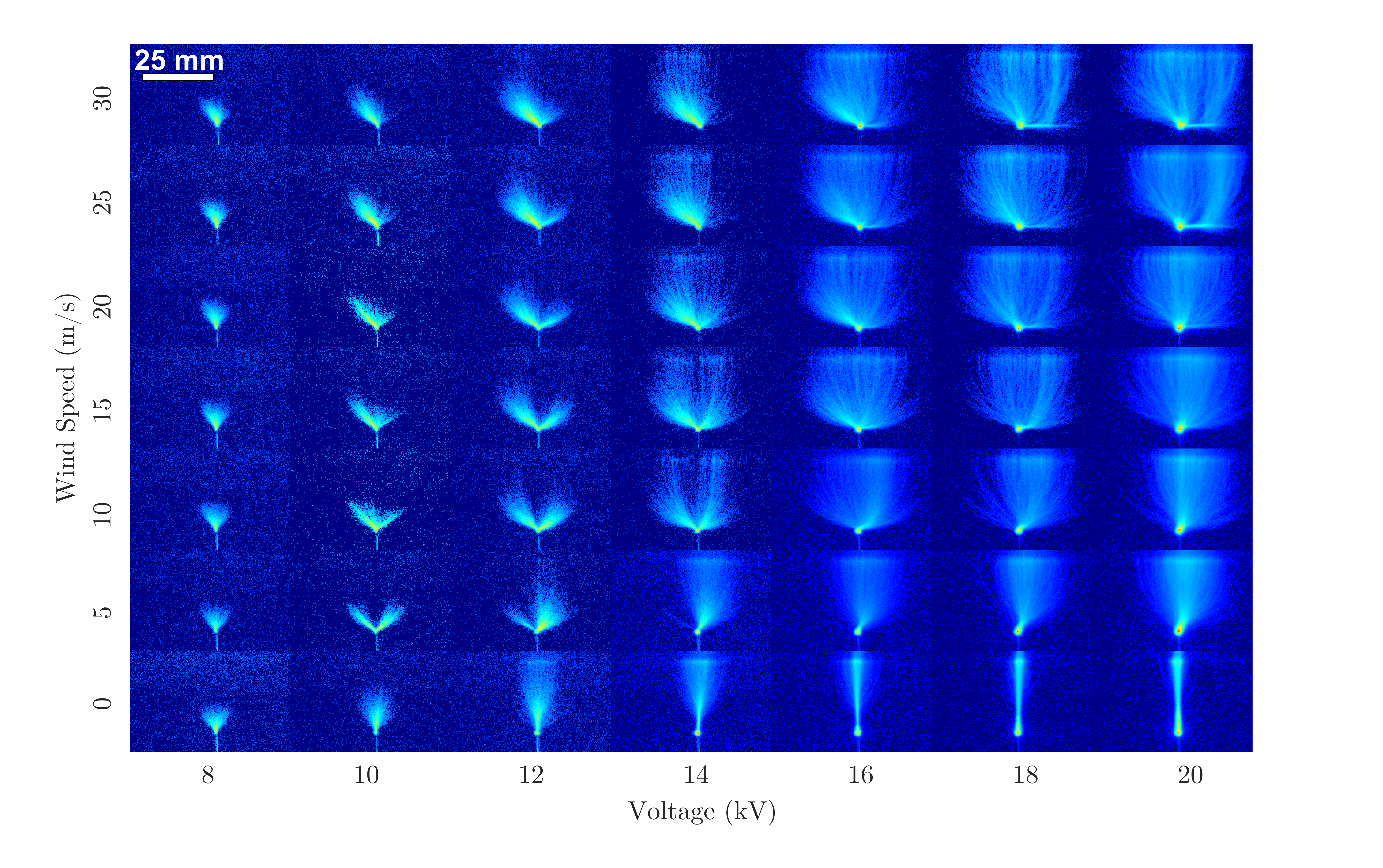}
    \caption{Time-integrated morphology of the discharge as a function of applied voltage and wind speed. Each image corresponds to the superposition of at least 45 streamer bursts. Wind going to the right.}
    \label{fig:allSuperimposed}
\end{figure*}

Figure~\ref{fig:allSuperimposed} shows the morphology of the discharge, as a function of applied voltage and wind speed, through the superposition of multiple ICCD images of single streamer bursts.

In the absence of wind, and at the lower applied voltages (6-10 kV), the discharge is a partial discharge, i.e., a streamer corona that does not bridge the gap. As the applied voltage is increased, the streamer corona extends further until it bridges the gap. Further increasing the applied voltage narrows the branching structure of the corona until it visually adopts a single filament morphology, Figure \ref{fig:allSuperimposed}. An increase in the applied voltage is also accompanied by a decrease in the period between pulsations (higher frequency), while the peak current and the energy deposition increase approximately quadratically, and the pulse width presents a narrow range of variation and peaks at about 14 kV. It is interesting to note that, at 10 kV, there is an anomaly in the observed frequency of pulsation of the streamer bursts, which goes from kHz to tens of kHz. This case coincides with the streamer corona transition to fully bridging the gap, Figure \ref{fig:allSuperimposed}.

The study of wind effects reveals novel observations compared to prior works \cite{Vogel2018ExperimentalFlow,niknezhad_three-dimensional_2021}. The work by Vogel et al. \cite{Vogel2018ExperimentalFlow} and Niknezhad et al. \cite{niknezhad_three-dimensional_2021} reports on streamers tilting in the direction of wind. In our experiments, this trend is observed at the higher voltages tested (14-20 kV) and lower wind speeds (0-10 m/s). In these cases, the streamer coronas are clearly directed in the downwind streamwise direction, Figure~\ref{fig:allSuperimposed}. At lower voltages (8-12 kV) and low wind speeds (0-10 m/s), the wind seems to have the effect of splitting and spreading out the direction of the streamers, both in the upwind and downwind directions, but favoring the downwind case. As the wind speed is increased further (15-30 m/s), more streamers are observed in the upwind direction, contrary to prior observations. For the higher voltages tested (14-20 kV) a similar effect is observed: at low wind speeds, streamers tend to propagate in the direction of the wind, but as the wind speed is increased, more and more streamers propagate against the wind. In addition, at the higher wind speeds tested, increased branching is observed as well as a broader range of streamer orientations. 

The increased dispersion in the streamer burst structure as the wind speed increases can also be appreciated in the electrical parameters of the discharge, Figures~\ref{fig:period_histograms} through \ref{fig:width_histograms}. The effect of wind for any given voltage can be visualized by focusing on each of the voltages (color-coded and labeled on the x-axis) separately. The period, peak current, energy deposition, and current pulse width, all show a tendency for increased dispersion with wind speed. In terms of general trends with wind speed, the inter-pulse period decreases with wind speed, Figure~\ref{fig:period_histograms}, and more of the high-frequency bursts appear, of the order of 100 kHz. As the wind speed increases, the peak current (Figure~\ref{fig:peak_histograms}) and energy deposition by individual pulsations (Figure~\ref{fig:energy_histograms}) both tend to increase dispersion, and at high voltages they decrease. The width of the current pulses (Figure~\ref{fig:width_histograms}) shows less variation with wind speed, other than a wider dispersion in its value. 

\subsection{Correlations between electrical properties and discharge morphology} \label{section:correlations}

\begin{figure}[htbp] 
    \centering
    \subfloat[10 kV, 0 m/s.\label{fig:3Superimposed_10-00}]{
        \includegraphics[width=.23\textwidth,trim={0 0 0 0}, clip]{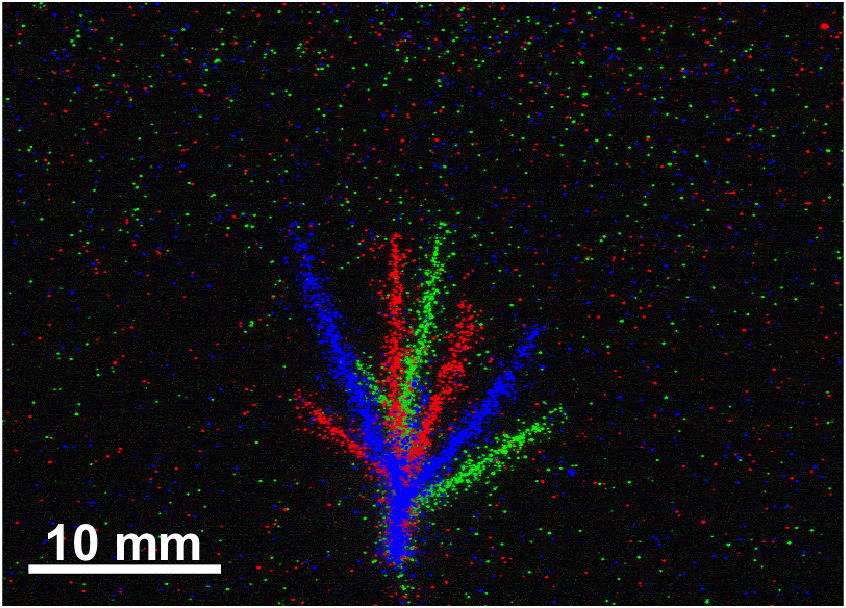}
    }
    \subfloat[10 kV, 30 m/s.\label{fig:3Superimposed_10-30}]{
        \includegraphics[width=.23\textwidth,trim={0 0 0 0}, clip]{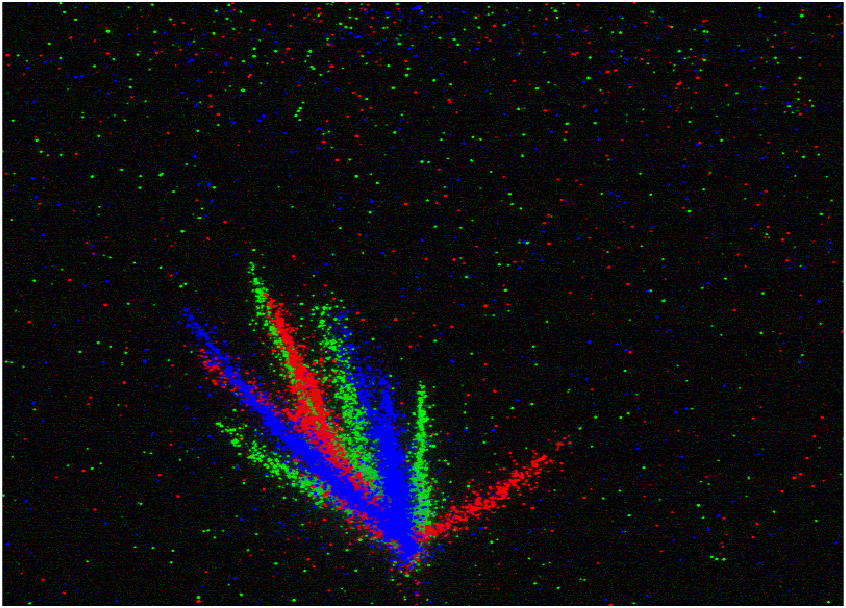}
    }
    
    \centering
    \subfloat[18 kV, 0 m/s.\label{fig:3Superimposed_18-00}]{
        \includegraphics[width=.23\textwidth,trim={0 0 0 0}, clip]{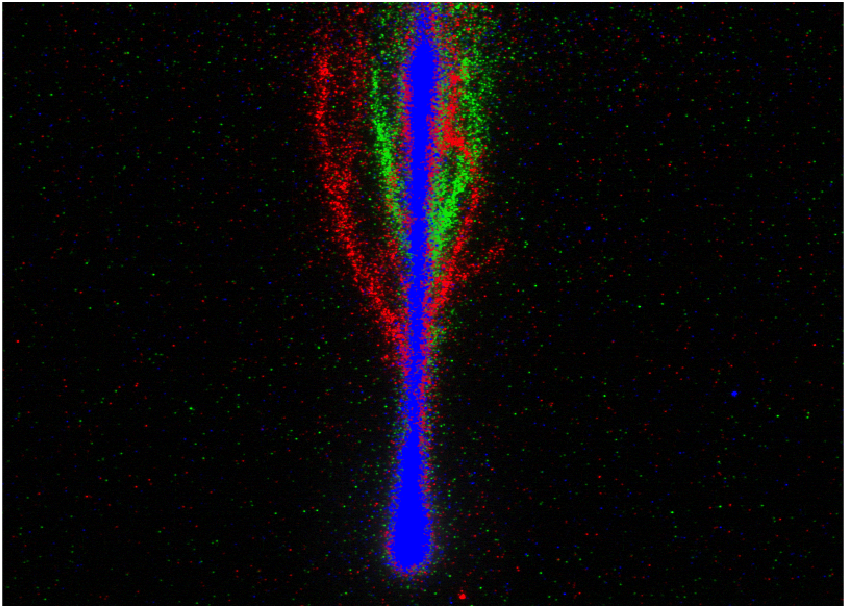}
    }
    \subfloat[18 kV, 30 m/s.\label{fig:3Superimposed_18-30}]{
        \includegraphics[width=.23\textwidth,trim={0 0 0 0}, clip]{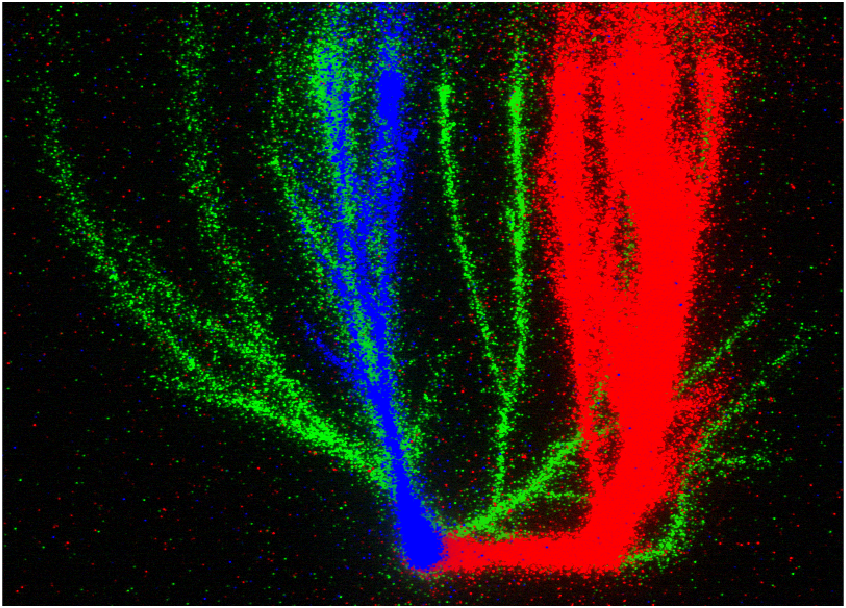}
    } 
    \caption{ICCD images showing three individual non-sequential streamer bursts, in different colors for two voltages and wind speeds. Wind is going to the right.}
    \label{fig:3Superimposed}
\end{figure}

The high dispersion in the electrical properties observed as the wind speed increases is also accompanied by a tendency for streamers to propagate in a wider range of orientations, including both downwind and upwind. For two voltages and two wind speeds, Figure \ref{fig:3Superimposed} shows three individual, non-sequential streamer bursts, in red, green, and blue. The images demonstrate that, within a fixed voltage and wind speed, individual streamer bursts can have different morphology and a tendency to propagate in different directions. The question now arises whether streamer bursts propagating in these different orientations can be correlated to disparate electrical properties, since particularly in the high wind speed cases, bi-modal and even multi-modal distribution functions are observed. 

\begin{figure*}[htbp] 
    \centering
    {\includegraphics[width=\textwidth,trim={0 0 0 0}, clip]{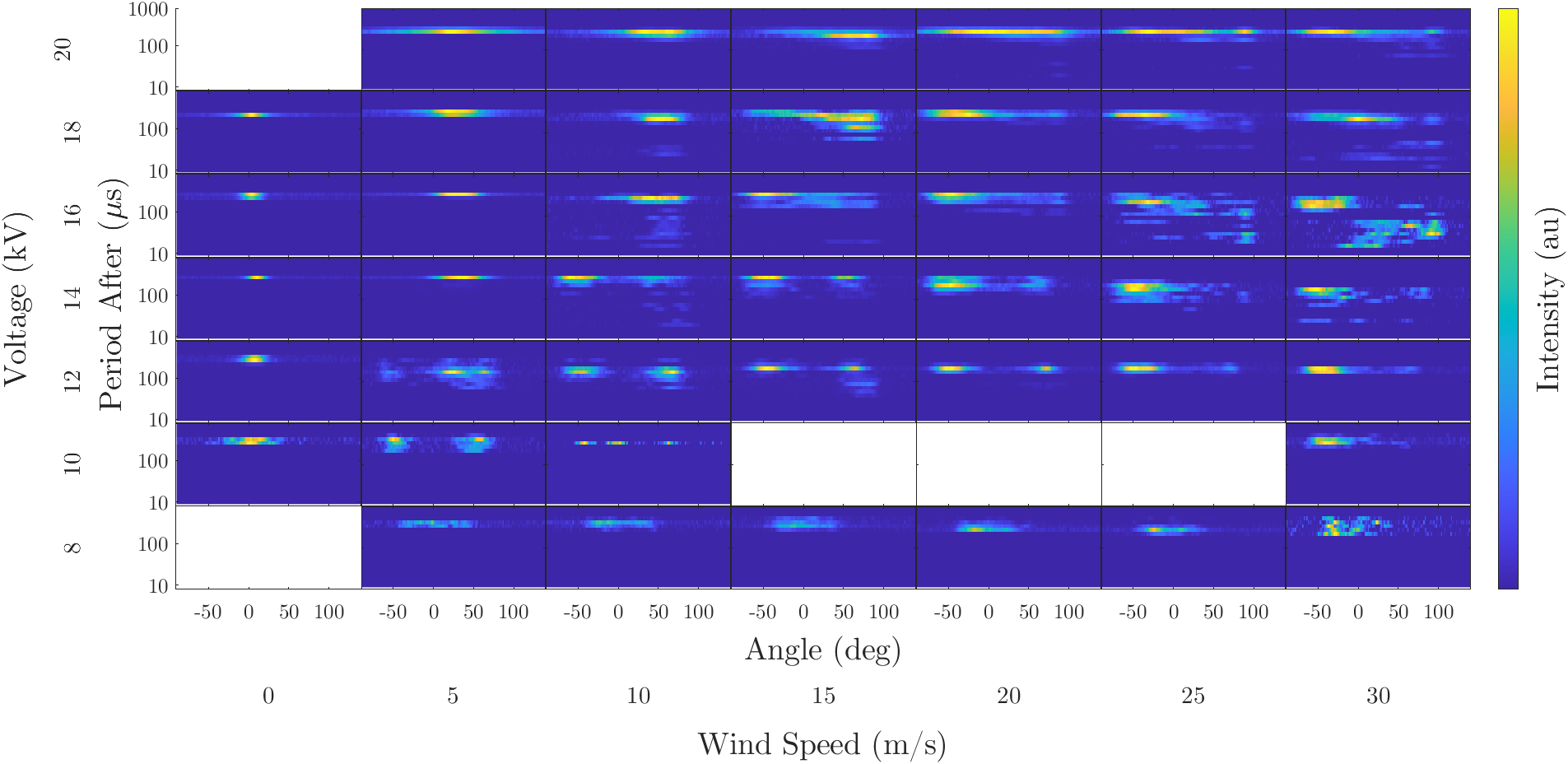}
    } 
    \caption{Correlation between streamer orientation, relative to the vertical with positive angles indicating the downwind direction (see Figure~\ref{fig_DiagnosticsExample}(a)), and inter-pulse period for all voltages and wind speed cases considered.}\label{fig:period_after_correlations}
\end{figure*}

\begin{figure*}[htbp]    
     \centering
    {\includegraphics[width=\textwidth,trim={0 0 0 0}, clip]{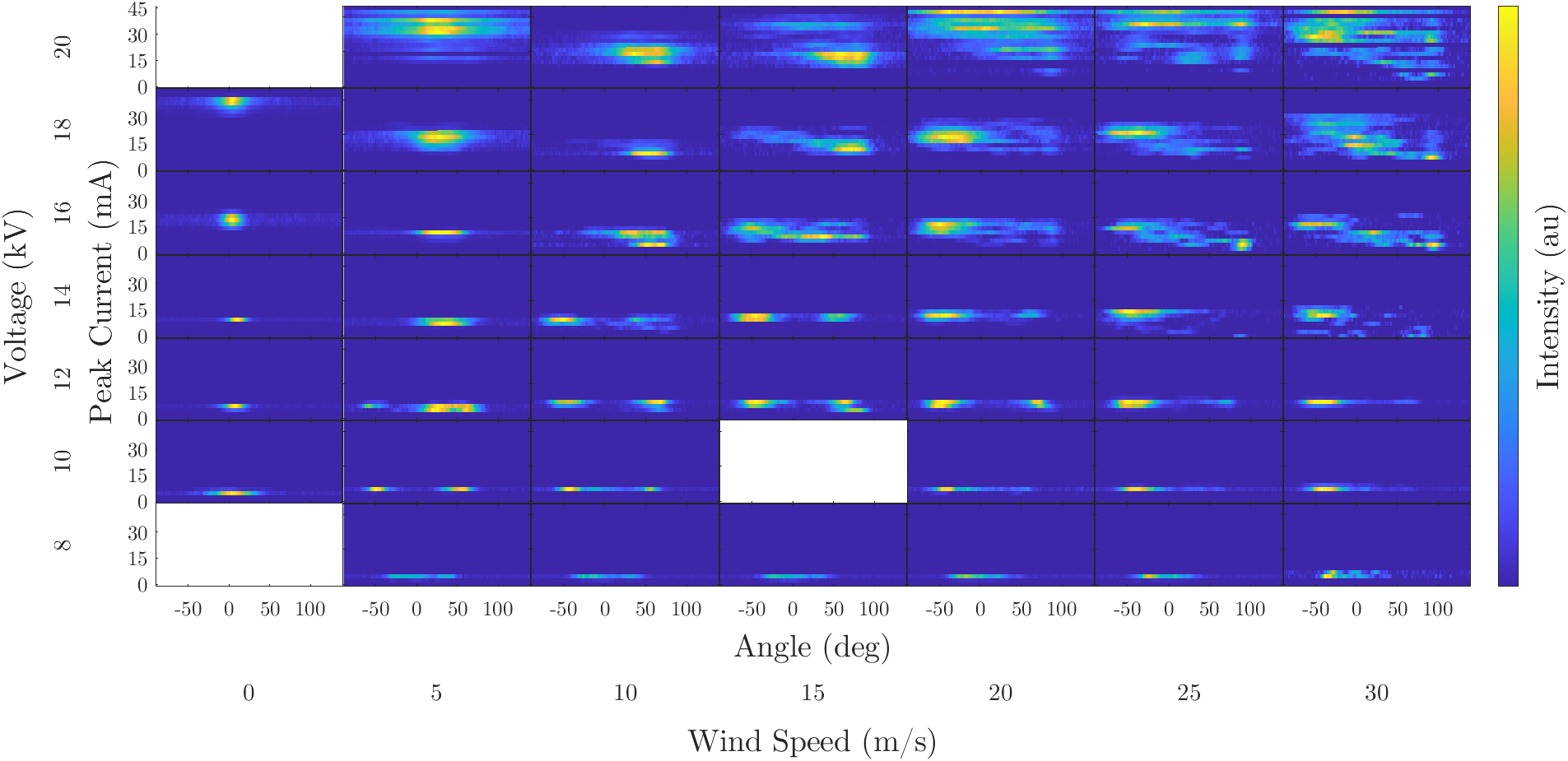}
    } 
    \caption{Correlation between streamer orientation, relative to the vertical with positive angles indicating the downwind direction (see Figure~\ref{fig_DiagnosticsExample}(a)), and peak pulse current for all voltages and wind speed cases considered.}\label{fig:peak_correlations}
\end{figure*}

To address this question, the methods described in section~\ref{section:synchronized} are used specifically to investigate the relationship between the streamer burst orientation (relative to the wind) and the pulsation period and peak current magnitude. The orientation of the streamer burst is quantified by taking the radially averaged intensity of each image, using polar coordinates and within a region of 0.5 mm about 5 mm away from the needle tip, as shown in Figure~\ref{fig_DiagnosticsExample} and as described by Guo et al \cite{guo_effect_2021}. Each image is then reduced to an angle-intensity profile which is paired to the corresponding inter-pulse period and peak current. All the images are then binned based on the inter-pulse period (measured from the pulse being imaged to the next) or peak current to reveal any correlation between streamer burst orientation and electrical properties. The period after the pulse being imaged is chosen (rather than the period before) as it is less affected by the biases described in section~\ref{section:synchronized}. This is plotted in Figures~\ref{fig:period_after_correlations} and \ref{fig:peak_correlations} for each wind speed and voltage case. Note that some cases (shown as blank) had a misalignment of the camera timing with the streamer burst, resulting in many images that captured more than one streamer burst or that missed the streamer burst in time, neither of which can be used for this analysis. 

Several observations can be made from these plots. First, for the no-wind cases, the streamer bursts are centered around the zero angle, as expected, and present a distinct characteristic value in both the inter-pulse period and peak current with some dispersion. This is consistent with the distribution functions in Figures~\ref{fig:period_histograms}-\ref{fig:width_histograms} presenting a single peak (normal distribution functions). As the wind speed is added, the streamers tilt in the direction of the wind (positive angles), and, for certain lower voltages (i.e. 10-12 kV) the corona splits into two directions with approximately half of the streamers biased downwind (positive angles) and half upwind (negative angles). The split streamers, upwind and downwind propagating respectively, present no difference in peak current magnitude or inter-pulse period. Moving to higher voltages and wind speeds, the streamers take a wider range of angles, as shown in Figure \ref{fig:3Superimposed_18-30}, with some streamer bursts propagating against the wind, some with the wind, and some covering multiple orientations. Many streamer bursts, such as the red colored in Figure \ref{fig:3Superimposed_18-30}, begin to propagate at a near 90 degree angle before turning and propagating towards the ground plate. In the high voltage and high wind speed cases, a dispersion in the current and period is also observed, that seems to correlate somewhat with streamer burst orientation. At the higher voltage and wind speeds, higher peak current streamers correspond with an upwind propagating burst, and low current bursts tend to propagate downwind. Similarly, after a streamer burst propagating in the upwind direction (higher peak current pulses), there is typically a longer inter-pulse period before the next streamer burst occurs. 

\begin{figure}[htbp]
        \includegraphics[width=.5\textwidth,trim={0 0 0 0}, clip]{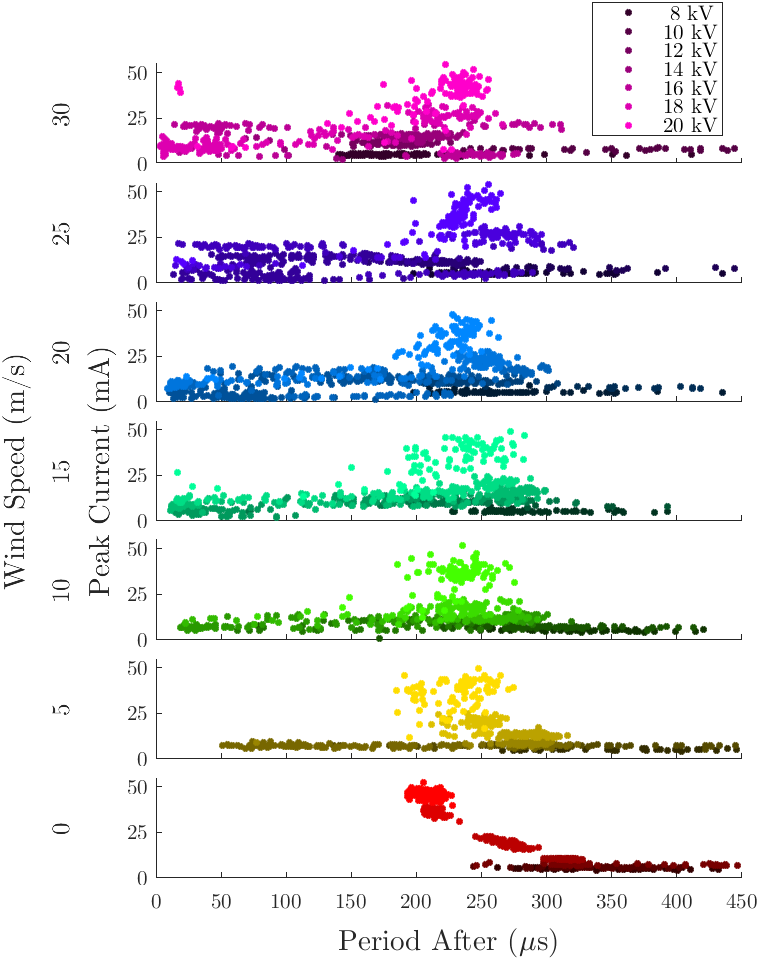}
    \caption{Correlation between streamer burst peak current and inter-pulse period (measured after the pulse under consideration) for all voltages and wind speeds tested.}
    \label{fig:periodCurrent}
\end{figure}

The relationship between the inter-pulse period and the peak current is plotted in Figure \ref{fig:periodCurrent}. At high wind speeds, the trends are consistent with the correlations observed in Figures \ref{fig:period_after_correlations} and \ref{fig:peak_correlations}. Higher current streamer burst seem to be associated with longer inter-pulse periods (after the burst), and low current streamers tend to have a shorter inter-pulse period afterwards. At low wind speeds, the dispersion is low, and lower voltages present lower peak currents and longer inter-pulse periods. 

\section{Discussion}

\subsection{Transient corona under DC voltage}

The self-pulsating streamer corona under DC voltage has been reported in many classical works \cite{Loeb1965,Goldman1978}. The results of this work are novel in that they measure the properties of individual pulsations in terms of distribution functions, and then account for possible airflow. 

In the transient corona regime of this work, the discharge self-pulsates in branching streamer bursts despite the constant voltage applied \cite{trinh_partial_1995, chang_corona_1991,bruggeman_foundations_2017,machala_dc_2008}. A streamer corona burst is quenched either by its own space charge that locally shields the electric field, or is limited by the current from the power supply and the stray capacitance and resistance of the circuit. In the absence of wind, the literature on coronas attributes the space charge removal, and electric field recovery, to be driven by the ion drift velocity \cite{Loeb1965}:
\begin{equation} 
u_{ion}=\mu E,
\end{equation}
where $\mu$ is the ion mobility and E is the electric field. In atmospheric pressure air and for positive polarity, $\mu \sim 1.4$ cm$^2$/V/s \cite{zhang_dependence_2017}, so for an average field of 6.5 kV/cm (comparable to that used in this work) the ion velocity is 90 m/s which leads to pulse repetition frequencies around $\sim 3.6$ kHz, for the charge to be evacuated from the 25 mm gap. This estimate is consistent with the measurements of this paper and similar frequencies of pulsation have been reported by many authors \cite{moreau_ionic_2018,arcanjo_observations_2021}. Based on this mechanism of charge removal, a higher applied voltage would lead to faster repetition frequencies, or shorter inter-pulse periods, as measured in this work. A more detailed estimate is given in \ref{appendix}, that correctly predicts the experimental trends with applied voltage and wind speed.

For no-wind, the peak pulse current increases approximately quadratically with applied voltage. This result might seem obvious based on prior knowledge of both glow and streamer corona \cite{townsend_electricity_1915_russian, arcanjo_observations_2021,guerragarcia_corona_2020}, however, it is not since normally constant DC current values or time-averaged values are reported. In this case, it is the \emph{peak} value of the current that shows this dependency, rather than a time-integrated measurement. 

\subsection{Impact of wind}

For streamer coronas in airflow, the transport of space charge will be enhanced by wind advection, so that the ion velocity is given by \cite{chapman_corona_1970,chapman_magnitude_1977,Nguyen2017,guerragarcia_corona_2020}:
\begin{equation}
    u_{ion}=\mu E+u_w,
\end{equation}
where $u_w$ is the local wind speed. For $\mu E >> u_w$, charge transport and the frequency of pulsation of the streamer bursts are dominated by the electrostatics. On the other hand, for $\mu E << u_w$, wind advection will accelerate ion evacuation, resulting in higher pulsation frequencies. In this work, the wind speeds tested are noticeable compared to a 90 m/s ion drift, which can explain the modest increase in the streamer burst pulsation frequencies with wind speed observed in Figure~\ref{fig:period_histograms} (see \ref{appendix}). However, this mechanism can not explain the observed high frequency streamer bursts that have frequencies of tens to hundreds of kHz.

The decrease in peak current with wind speed at high voltages might seem counter-intuitive at first. Previous studies of corona in wind typically show a corona current that increases with wind speed, with a linear dependency for high enough winds compared to the electric drift \cite{chapman_corona_1970,chapman_magnitude_1977,Nguyen2017}. This dependency can be easily explained by the enhanced ion speed in wind. However, the majority of these studies focus on the glow corona regime \cite{chapman_corona_1970,chapman_magnitude_1977,guerragarcia_corona_2020} and the results correspond to DC current measurements. For the case reported in this paper, the regime is different (streamer corona) and the measurement is a peak value, not the time-integrated one. Note that, as the frequency of pulsation increases, the \emph{average} current does not necessarily decrease even if the \emph{peak} value does. In addition, while the 20 kV case shows a clear decrease in current, the lower voltages disperse to both higher and lower current bursts. 

As wind tilts the streamer corona, it typically increases its propagation length. For comparable conductivity and cross-section of the filaments, this increase in length could explain a decrease in current with higher wind speeds, since longer streamers would have a higher resistance. The red streamer burst in Figure \ref{fig:3Superimposed_18-30} is about 1.4 times longer than the blue in \ref{fig:3Superimposed_18-00}. 

It is also worth noting that, in these DC conditions, the self-pulsating streamers typically appear superimposed to a localized glow at the anode \cite{Goldman1978}. In that case, an intensified glow discharge with wind (with increasing ion densities with wind, as reported by \cite{Nguyen2017}) may result in lower current streamers as they develop in the space charge cloud of the glow. The presence of the glow may also be impacting the propagation direction of the streamers. As the glow is intensified, it provides more space charge shielding in the downwind side of the electrode \cite{Nguyen2017}, opening up new regions of streamer development at the upwind side. 

A similar effect can be attributed to the local velocity field around the tip: a turbulent recirculation region can locally trap positive ions, reducing the local electric field just downwind of the tip and making streamer inception and propagation more likely in the upwind direction. This flow recirculation region was modelled by Niknezhad et al. \cite{niknezhad_three-dimensional_2021}. While the exit flow from the wind tunnel is relatively laminar, the ground plate and the electrode tip can create turbulence with characteristic Reynolds numbers of more than $10^5$.  The turbulent boundary layer over the ground plate can be estimated to be $\sim$3 mm \cite{schlichting_boundary-layer_1979}, which is small but not insignificant compared to the 25 mm gap. The dynamic pressure of the flow is $\sim$550 Pa, for a 30 m/s wind speed. Since this value is small compared to the atmospheric pressure of 101 kPa, it is likely not a contributor to changes in the electrical properties of the discharge. However, it may be a contributor to opening up new streamer burst locations along the electrode in fast succession (high frequency streamer bursts) as close to the breakdown threshold, small modifications in the pressure could trigger or suppress a discharge.

All of these effects also contribute to augmenting the branching of the discharge: the wind disperses the residual charges, pre-ionizing regions of the flow-field and favoring the propagation of successive streamers. In the no-wind case, electrons and ions follow fairly vertical paths, which guide the propagation of the next streamer \cite{wu_positive_2018}. Local heating by the current passage also contributes to favoring a particular orientation, through increased reduced electric fields. This is more noticeable for discharges bridging the gap and single filament regimes.

Finally, the results are also likely affected by the circuit properties which remain unchanged between no-wind and wind cases.

\section{Conclusions}
This work presented novel methods of characterizing self-pulsating DC discharges based on statistical analysis, and reported on the effect of applied voltage and wind speed on the streamer corona regime. The study focuses on the impact of these parameters on the distribution functions of inter-pulse period, peak current amplitude, energy deposition per pulse, and pulse width. In general, the electrical parameters presented more dispersion as the wind speed increased. The mean of the inter-pulse period decreases
with wind speed and the mean pulsation frequency increases. The peak currents and energies per pulsation have a general tendency to decrease in magnitude as wind speed increases; but also higher-current, higher-energy, streamer bursts are observed. Streamer busts with very high frequencies (order 100 kHz) were captured that can not be explained by ion evacuation enhancement due to the presence of wind. In the presence of wind more of these very high frequency bursts are observed. At low wind speeds (0-10 m/s), the streamer bursts tend to propagate with the direction of the wind, as reported in other works. However, as the wind speed was increased (in the 15-30 m/s range), streamers propagating against the wind appeared, with a wide dispersion in propagation direction observed. Finally, the streamer burst orientation with respect to wind was correlated with the electrical properties measured, through synchronized imaging. These measurements revealed that upwind streamers tend to have a higher peak current and longer inter-pulse period before the next burst appears.

\section*{Acknowledgments}
This  work  was partially funded  by  The  Boeing  Company  through the Strategic Universities for Boeing Research and Technology Program. L. Strobel acknowledges support through a Mathworks fellowship. The authors would like to thank Claire Johnson and Thomas Edelman for testing support. 

\appendix
\section{Pulsation frequency driven by charge transport}\label{appendix}

The influence of applied voltage and wind on the inter-pulse period (and pulse repetition frequency) is here estimated using an analytical model of charge transport. The electrostatics of the needle-to-plate electrode setup is approximated by a hyperboloid of revolution with the same radius of curvature at the tip as the needle (100 $\mu$m) and the same distance to the plate as in the experiment (25 mm). The solution to Laplace's equation is obtained using prolate spheroidal coordinates to solve for the electric fields in the discharge gap prior to any space charge being injected. The boundary of the corona zone is estimated using the classical assumption that the field inside the corona is equal to the stability value (e.g., 500 kV/cm for the positive polarity in atmospheric pressure air \cite{Gallimberti2002,Gallimberti1979}). Using this approximation, the corona boundary is calculated as the envelope where the electrostatic potential in the absence of space charge matches the electrostatic potential calculated assuming that the field is radial from the tip with magnitude equal the stability value \cite{Arevalo2012}. From the estimated electric fields in the discharge gap, the drift velocity field of the ions can be calculated using a positive ion mobility of $\mu \sim 1.4$ cm$^2$/V/s \cite{McDaniel1973}. In the absence of wind, the ion trajectories will follow the electric field lines and a characteristic time for ion removal can be calculated as the time it takes for an ion to move from the tip to the grounded plate. When accounting for wind, the trajectories of the ions are modified by superimposing a uniform wind velocity, orthogonal to the tip, to the electric drift. 

\begin{figure}[htbp]
        \includegraphics[width=.5\textwidth,trim={200 100 200 200}, clip]{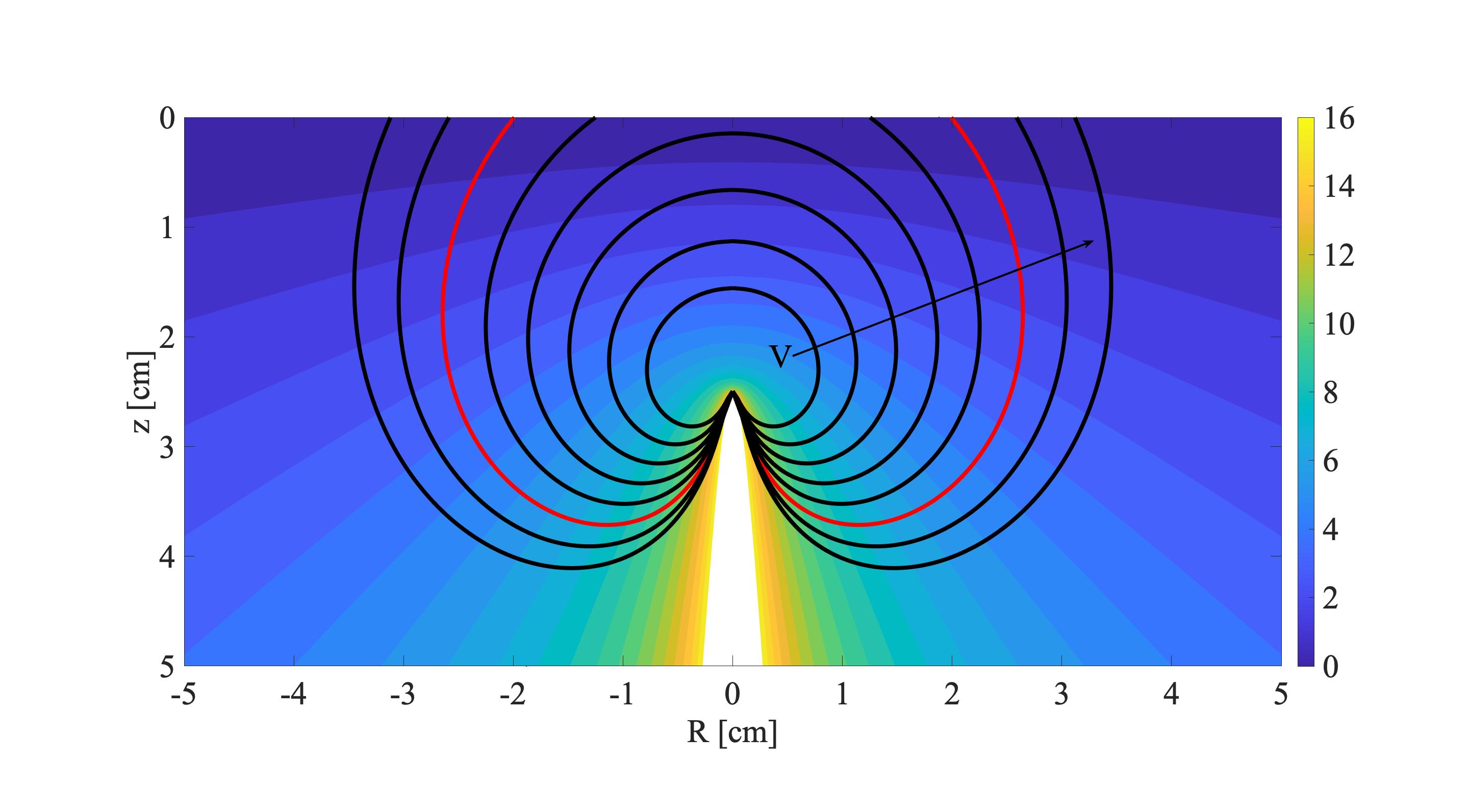}
    \caption{Electrostatic potential calculation in kV for an applied voltage of 16 kV. The calculated corona boundary is marked in red. Corona boundaries for voltages between 6 kV and 20 kV in 2 kV increments (as in the experiment) are marked in black.}
    \label{fig:appendix}
\end{figure}

An example of the calculated electrostatic potential and corona boundary is shown in Figure~\ref{fig:appendix}, for the case of 16 kV. The plot also shows the calculated corona boundaries for voltages between 6 kV and 20 kV in 2 kV increments (as in the experiment). The model predicts that the corona first bridges the gap between 12 and 14 kV, as in the experiment, Figure~\ref{fig:allSuperimposed}. 

\begin{figure}[htbp]
        \includegraphics[width=.5\textwidth,trim={200 100 200 200}, clip]{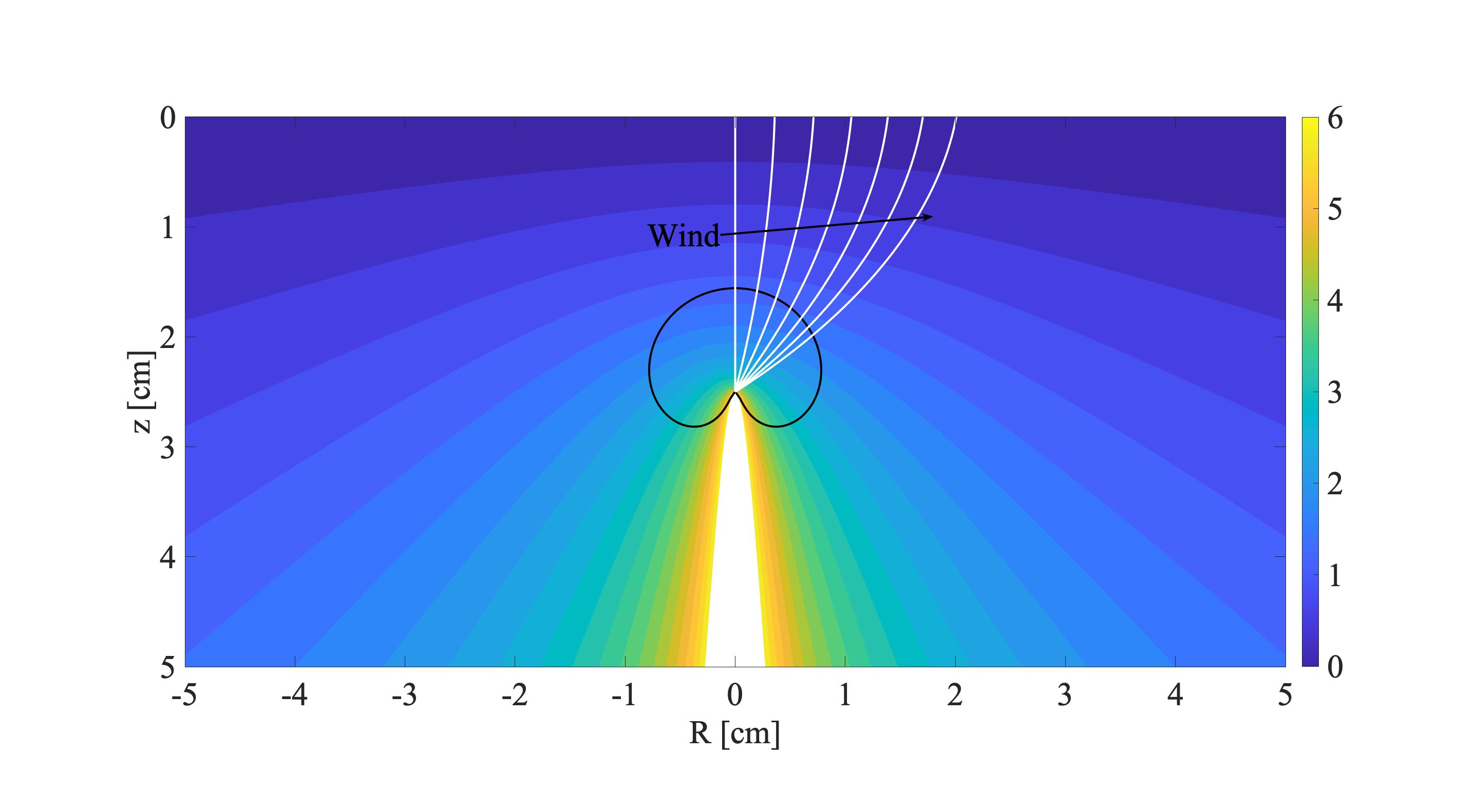}
    \caption{Electrostatic potential calculation in kV for an applied voltage of 6 kV. The calculated corona boundary is marked in black. The calculated ion trajectories in the presence of wind are marked in white for wind speeds between 0 m/s and 30 m/s in 5 m/s increments (as in the experiment).}
    \label{fig:appendix4}
\end{figure}

Figure~\ref{fig:appendix4} shows the calculated electrostatic potential and corona boundary for the case of 6 kV with the calculated ion trajectories in the presence of wind marked in white. In this configuration, the wind-to-electric drift ratio is the highest of all cases tested so the deviation of the ion trajectories from the electric field lines should be the greatest. Even in this situation, the trajectories are well contained within the 203 mm diameter grounded plate, so that little to no leakage of the current in the direction of the wind is expected. This is consistent with the measurements taken by the two current probes at the high voltage and grounded side respectively, which show very similar results.

\begin{figure}[htbp]
        \includegraphics[width=.5\textwidth,trim={0 0 0 0}, clip]{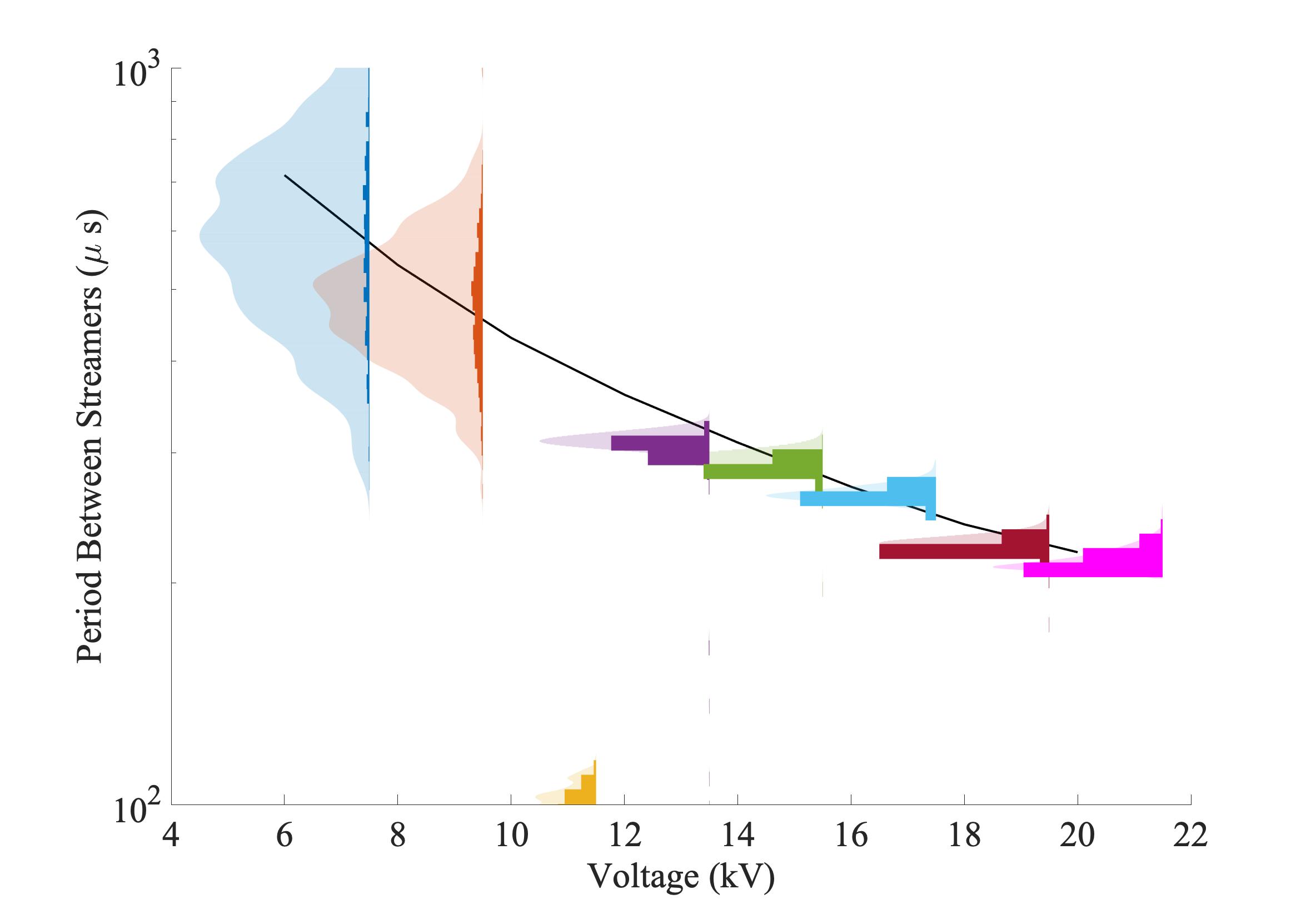}
    \caption{Calculated period between streamer bursts (black line) superimposed to experimental data as a function of applied voltage. Case of no wind.}
    \label{fig:appendix2}
\end{figure}

Using these approximations, the effect of increasing the applied voltage (no-wind case) is shown in Figure~\ref{fig:appendix2}. The trend and magnitude of the inter-pulse periods are well predicted by a simple model of charge transport in the Laplacian electric field solution.

\begin{figure}[htbp]
        \includegraphics[width=.5\textwidth,trim={0 0 0 0}, clip]{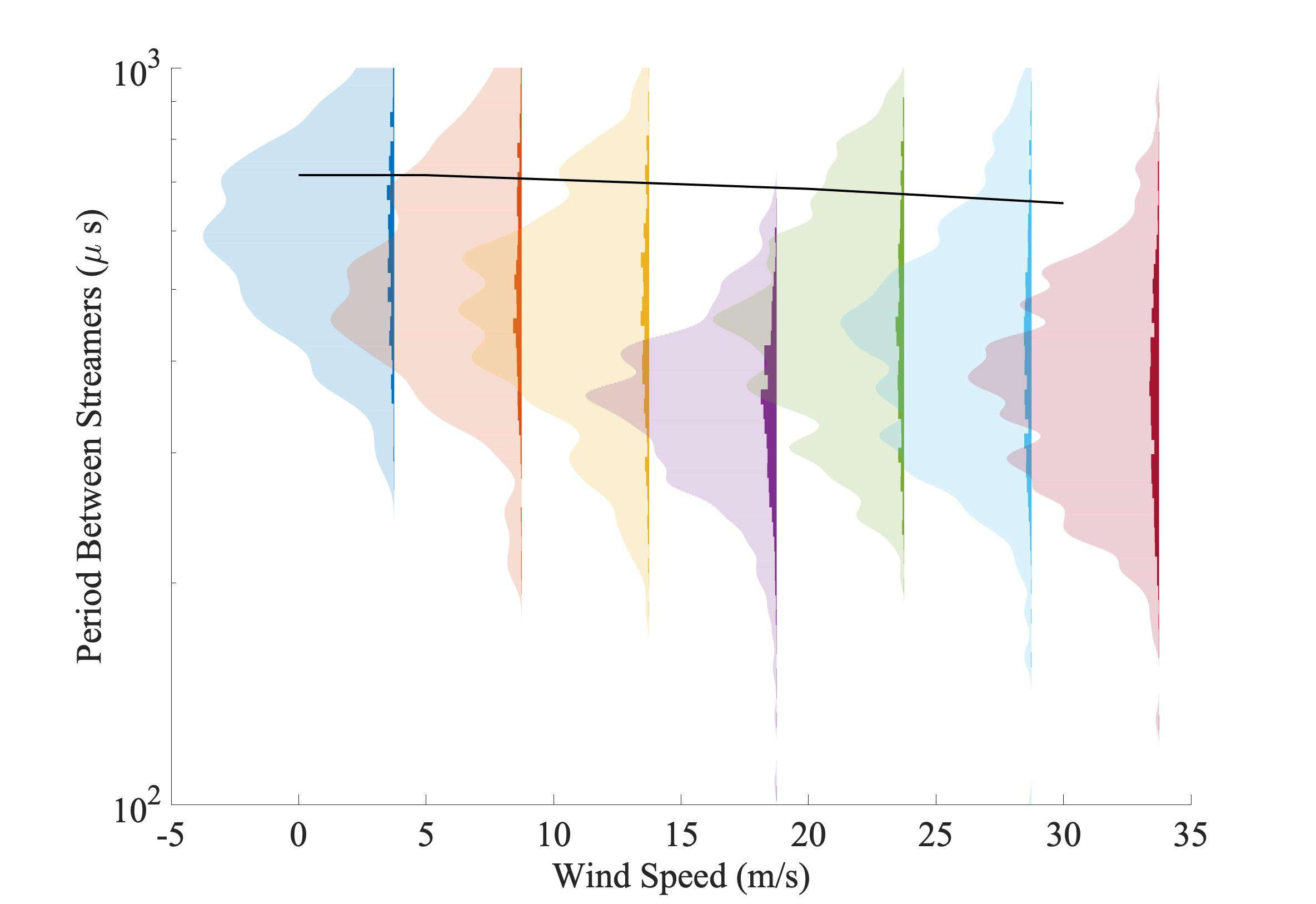}
    \caption{Calculated period between streamer bursts (black line) superimposed to experimental data as a function of wind. Case of 6 kV.}
    \label{fig:appendix3}
\end{figure}

A similar calculation is performed fixing the applied voltage to 6 kV and varying the wind speed in the experimental range of 0 to 30 m/s. The low voltage case is selected because of the absence of the very high frequency streamer bursts that can not be addressed by charge transport. Also, the effect of wind (compared to electric drift) should be more noticeable in this case, compared to the higher applied voltage cases. The results are shown in Figure~\ref{fig:appendix3}, superimposed to the experimental data. The decreasing period (increasing frequency) with wind speed is predicted by the model although the modification is modest and the magnitude is less than measured experimentally. The method used to estimate the period might be under-predicting the wind effect: the speed is higher but the trajectory to reach the ground plate is longer, Figure~\ref{fig:appendix4}.


\section*{References}
\bibliographystyle{iopart-num}
\bibliography{references.bib,references_BCM.bib, references_BCM_Mendeley.bib,references_CGG}

\end{document}